\documentclass[12pt,a4paper]{article}

\newcommand{\pa}{\partial}
\newcommand{\al}{\alpha}
\newcommand{\be}{\beta}
\newcommand{\ga}{\gamma}
\newcommand{\de}{\delta}

\begin{document}

\begin{flushright}
{ }
\end{flushright}
\vspace{1.8cm}

\begin{center}
 \textbf{\Large Conformal Transformations and Strings for\\
an Accelerating Quark-Antiquark Pair in $AdS_3$}
\end{center}
\vspace{1.6cm}
\begin{center}
 Shijong Ryang
\end{center}

\begin{center}
\textit{Department of Physics \\ Kyoto Prefectural University of 
Medicine \\ Inamori Memorial Hall \\
Shimogamo, Kyoto 606-0823, Japan}
\par
\texttt{ryang@koto.kpu-m.ac.jp}
\end{center}
\vspace{2.8cm}
\begin{abstract}
From a simple moving open string solution  dual to a moving heavy quark
with constant velocity in the Poincare $AdS_3$ spacetime, we construct 
an accerlerating open string solution dual to a heavy quark-antiquark 
pair accelerated in opposite directions by performing the three mappings
such as the  $SL(2,R)_L \times SL(2,R)_R$ isometry transformation, 
the special conformal transformation and the conformal SO(2,2) 
transformation. Using the string sigma model action we construct
two open string solutions staying in two different regions whose
dividing line is associated with the event horizon appeared on the
 string worldsheet and obtain the accelerating open string solution
by gluing two such solutions.
\end{abstract}
\vspace{3cm}
\begin{flushleft}
December, 2014
\end{flushleft}

\newpage
\section{Introduction}

The AdS/CFT correspondence \cite{JM} has more and more revealed 
the strong coupling behaviors of the $\mathcal{N}=4$ super 
Yang-Mills (SYM) theory by using the string theory in $AdS_5 \times S^5$
where various open and closed string solutions are studied. 

By using the Nambu-Goto action there have been constructions of the moving
open string solutions with 
constant velocity in $AdS_5$ black hole 
geometries \cite{HKY,SG}, where the dynamics of quark moving in strongly
coupled $\mathcal{N}=4$ SYM thermal plasma is investigated by regarding
the infinitely massive quark as the open string end at the boundary of
the $AdS_5$-Schwarzschild spacetime \cite{SG} and the finitely 
massive quark as the open string end at the D7-brane \cite{KK} in the
$AdS_5$-Schwarzschild spacetime \cite{HKY}.
 
Mikhailov has used the Nambu-Goto action 
in the Poincare $AdS_5$ spacetime to
present an analytic generic solution for the open string dual to a single
infinitely massive quark moving on an arbitrary timelike trajectory
in the $\mathcal{N}=4$ SYM theory and extract a rate of the energy loss
which agrees with the Lienard formula \cite{AM}.
Based on the extension of this generic solution to the finite quark mass
case it has been shown that an event  horizon 
appears on the worldsheet whenever
the single quark accelerates in any fashion \cite{CG}.
The division of an open string  through the horizon is associated with 
two contributions to the energy-momentum of the string corresponding to
the intrisic and radiated energy-momentum of the quark \cite{CG,CGG}.

An accelerating open string solution dual to a heavy quark-antiquark
pair uniformly accelerated in opposite directions has been found \cite{BX}
by using the Nambu-Goto action in the Poincare $AdS_5$ spacetime. 
The event horizon has been shown to appear on the worldsheet of the
open string connecting a quark and an antiquark and separate the radiation
 and the quark. It has been demonstrated that the rate of energy flow 
across the horizon becomes the same as derived from the Lienard 
formula of ref. \cite{AM}.

The similar accelerating open string solution has been constructed by
analyzing the Nambu-Goto action in the Rindler spacetime which is 
given by a coordinate transformation from the AdS spacetime \cite{PPZ}.
It has been studied that the energy loss via the moving open string
in $AdS_5$-Schwarzschild spacetime is related with the appearance
of the worldsheet horizon \cite{SSG,BMX} and the worldsheet Hawking
radiation generates the stochastic motion of the quark \cite{BHR}.

The accelerating string solution \cite{BX} associated with
a uniformly accelerating quark-antiquark pair has been constructed 
\cite{CCG,JAP} as a particular instance
from the generic string solution  \cite{AM} 
 dual to a single quark moving on an
arbitrary trajectory. The generic string solution has been indirectly
shown \cite{AM} to extremize the action, and further has
been directly substituted into the string equation of motion and 
confirmed to solve it \cite{GP}.
There have been various investigations of the thermal effects of the
worldsheet horizon on the accelerating string which are associated with
the Unruh temperature \cite{CCG,CP,HKK}.  

Starting from the generic string 
solution in the Poincare $AdS_5$ spacetime
\cite{AM} and using a suitable coordinate transformation, the accelerating
open string solution dual to a single accelerating quark in the global 
$AdS_5$ spacetime has been constructed \cite{HS} (see also \cite{FGP}). 

Recently it has been conjectured that the entanglement of
the general quantum Einstein-Podolsky-Rosen (EPR) pair 
 is intimately related with the 
Einstein-Rosen bridge or the non-traversable wormhole \cite{MS}.
Associated with the existence of horizon on the worldsheet of the 
accelerating open string dual to a uniformly 
accelerating quark-antiquark pair,
there have been several studies where the quark-antiquark pair is 
concretely regarded as a color singlet EPR pair in the $\mathcal{N}=4$ 
SYM theory and its entanglement is encoded in a non-traversable
wormhole on the worldsheet of the flux tube connecting the pair 
\cite{JK,JS,CGP,JKR}. The entanglement 
entropy of the quark-antiquark pair
has been investigated \cite{JK,LM,VHS} and the relation between the 
entanglement entropy and the string surface describing gluon 
scattering in position space has been studied \cite{SS}.

We will use the Nambu-Goto action in the static gauge for the open string
in the Poincare $AdS_3$ spacetime and  make an ansatz for the string 
profile expressed by three 
parameters to reconstruct the two string solutions
associated with the one-cusp Wilson loop \cite{MK} and the accelerating
string solution dual to an accelerating quark-antiquark pair \cite{BX}.

We will consider the $SL(2,R)_L \times SL(2,R)_R$ 
isometry group of the $AdS_3$ spacetime \cite{JMS,BHT} and make this 
isometry transformation for a simple
moving string solution  dual to a moving quark with constant velocity 
to construct the  accelerating string solution dual to a uniformly
accelerating quark-antiquark pair. This isomery transformation will be 
applied further to the accelerating string solution.  For the moving
 string solution with constant velocity and the  
accelerating string solution  we will perform the special
conformal transformation in the Poincare coordinates
and the conformal SO(2,2) transformation
in the embedding coordinates to see what kinds of string solutions appear.

Based on the string sigma model action we will make a special ansatz
for the string profile in the factorized form 
and construct the accelerating string solution on which the event
horizon appears.

\section{The accelerating string in the Nambu-Goto action}

Based on the Nambu-Goto action we consider an open string in 
$AdS_3$ with the Poincare metric
\begin{equation}
ds^2 = \frac{dz^2 - dt^2 + dx^2}{z^2},
\label{me}\end{equation}
where we have set the AdS radius $R$ to unity.
We use the static gauge
\begin{equation}
t = \tau, \hspace{1cm} z = \sigma
\label{st}\end{equation}
to express the string action in the Lorentzian worldsheet coordinates
\begin{equation}
S = - \frac{\sqrt{\lambda}}{2\pi}\int d\tau d\sigma \frac{\sqrt{D}}
{\sigma^2}
\end{equation}
with $D = 1 - (\pa_{\tau}x)^2 + (\pa_{\sigma}x)^2$.
In the equation of motion for $x$
\begin{equation}
\pa_{\tau}\left( \frac{\pa_{\tau}x}{\sigma^2 \sqrt{D}} \right) =
\pa_{\sigma}\left( \frac{\pa_{\sigma}x}{\sigma^2 \sqrt{D}} \right) 
\label{eq}\end{equation}
we make an ansatz
\begin{equation}
 x = \pm \sqrt{A\tau^2 + B - C\sigma^2}
\label{pa}\end{equation}
to have 
\begin{equation}
A\pa_{\tau}\left( \frac{\tau}{\sigma^2 \sqrt{F}} \right) = - C
\pa_{\sigma}\left( \frac{1}{\sigma \sqrt{F}} \right) 
\end{equation} 
with $F = A(1-A)\tau^2 + B - C(1-C)\sigma^2$.

For $B \neq 0$ there is one solution specified by $A = C = 1$
which yields the string configuration
expressed in terms of $B \equiv b^2$ as
\begin{equation}
x = \pm \sqrt{t^2 + b^2 - z^2}.
\label{ex}\end{equation}
This string solution was found in ref. \cite{BX} where the infinitely 
massive quark and antiquark are
located on the hyperbolic trajectories $x = \pm \sqrt{t^2 + b^2}$ at the
AdS boundary $z = 0$ such that the plus/minus sign of (\ref{ex})
represents the right  and left half of the accelerating string.
 The quark and antiquark first approach to each
other in decelerating and stop to return back in accelerating away
from each other with proper acceleration $1/b$.

For $B = 0$ there are two solutions which are provided by
$A\neq 0, 1$ with $C = 0$ and  $C \neq 0, 1$ with $A = 1$. 
The former gives a simple solution
\begin{equation}
x = \pm \sqrt{A}t,
\label{xts}\end{equation}
while the latter is complementary to (\ref{ex}) and leads to
a string solution
\begin{equation}
x = \pm \sqrt{2C\left(t^2 - \frac{z^2}{2}\right)},
\label{tzc}\end{equation}
whose $C$ is fixed as $C = A/2 = 1/2$.
The latter string configuration is  described by
\begin{equation}
z = \sqrt{2(t^2 - x^2)},
\end{equation}
which is the one-cusp Wilson loop solution of \cite{MK},
where the open string surface ends on two semi infinite
lightlike lines. This solution yields pure imaginary Lagrangian
so that the amplitude shows the exponential suppression.
The planar four-gluon scattering amplitude was computed by using 
the four-cusp Wilson loop solution in the T-dual AdS spacetime which was
obtained from the one-cusp Wilson loop solution by perfoming the conformal
SO(2,4) transformation \cite{LAM}. 

In \cite{SR} the following two-cusp Wilson loop solution 
was constructed by applying  the conformal SO(2,4) transformation to the
 one-cusp Wilson loop solution
\begin{equation}
z^2 = t^2 - x^2 \pm \sqrt{2(t^2 + x^2) - 1},
\label{zsq}\end{equation}
whose surface ends on four lines $t = x \pm 1, \; t = - x \pm 1$ which 
meet at two cusps $(t,x) = (0,\pm 1)$ for the plus sign and
 two cusps $(t,x) = (\pm 1,0)$ for the minus sign.
The appropriate square 
of (\ref{zsq}) yields an equation
for $x$ in the fourth order whose solution is given by
\begin{equation}
x^2 = t^2 + 1 - z^2 \pm \sqrt{4t^2 - 2z^2}.
\label{xtz}\end{equation}
As the string solution in the Poincare coodinates can be rescaled as
$x^{\mu} = (t,x) \rightarrow x^{\mu}/b, \; z \rightarrow z/b$, 
the expression (\ref{xtz}) becomes
\begin{equation}
x = \pm \left( t^2 + b^2 - z^2 \pm 2b\sqrt{t^2 - \frac{z^2}{2}} 
\right)^{1/2},
\label{bzt}\end{equation}
which shows a suggestive expression that contains two polynomials
$t^2 + b^2 - z^2$ and $t^2 - z^2/2$ of (\ref{ex}) and (\ref{tzc}).
The expression (\ref{bzt}) as a convolution of two square roots
is confirmed indeed to solve
the string equation of motion (\ref{eq}) through 
$D = -b^2\sigma^2/4x^2(\tau^2 - \sigma^2/2)$.

\section{Conformal transformations}

We use $w^{\pm} \equiv x \pm t$ to rewrite the Poincare metric (\ref{me})
as $ds^2 = (dz^2 + dw^+ dw^-)/z^2$. 
The $SL(2,R)_L \times SL(2,R)_R$ isometry group for this
$AdS_3$ metric was investigated in ref. \cite{JMS} (see also \cite{SRY}).
The $SL(2,R)_L$ transformation is given by
\begin{eqnarray} 
w^+ \rightarrow {w^+}' &=& \frac{\al w^+ + \be}{\ga w^+ + \de},
 \hspace{1cm}
w^- \rightarrow {w^-}' = w^- + \frac{\ga z^2}{\ga w^+ + \de}, \nonumber \\
z \rightarrow z' &=& \frac{z}{\ga w^+ + \de} 
\label{sl}\end{eqnarray}
with real $\al, \be, \ga, \de$ obeying $\al \de - \be \ga = 1$,
while the $SL(2,R)_R$ transformation is
\begin{eqnarray} 
w^+ \rightarrow {w^+}' &=& w^+ + \frac{\ga z^2}{\ga w^- + \de}, 
\hspace{1cm}
w^- \rightarrow {w^-}' = \frac{\al w^- + \be}{\ga w^- + \de}, 
 \nonumber \\
z \rightarrow z' &=& \frac{z}{\ga w^- + \de}. 
\label{slr}\end{eqnarray}
In view of (\ref{sl}) and (\ref{slr}) $\al$ and $\de$ have no dimension 
whereas $\be$ and $1/\ga$ have the same dimension as $w^{\pm}$.
Both transformations map the $AdS_3$ boundary to itself and act on the
boundary as the usual conformal transformations of (1+1)-dimensional
Minkowski spacetime. 

Let us consider a string configuration which extends straight
from the AdS boundary
at $z = 0$ to the Poincare horizon at $z = \infty$ 
in the $z$ direction and moves with constant
velocity $v$ in the $x$ direction
\begin{equation}
x = vt,\;\; t = \tau, \;\; z = \sigma,
\label{cm}\end{equation}
which simply satisfies the string equation (\ref{eq})
as is seen in (\ref{xts}),  and is dual to an
isolated infinitely-massive
quark moving with constant velocity $v$.
We perform the $SL(2,R)_L$ transformation with $\ga \neq 0$ for 
the straight moving string solution to obtain an accelerating 
string configuration associated with proper
acceleration $\ga\sqrt{(1+v)/(1-v)}$ of a quark-antiquark pair
in the following form
\begin{equation}
x' - \frac{1}{2\ga} \left( \al + \frac{1-v}{1+v} \de\right) = 
\pm \Bigg[ \left( t' - \frac{1}{2\ga} \left( \al - \frac{1-v}{1+v} \de
\right) \right)^2 + \frac{1-v}{1+v}\frac{1}{\ga^2} - {z'}^2 \Bigg]^{1/2},
\label{al}\end{equation}
where there are some constant shifts in $x'$ and $t'$ compared with
(\ref{ex}).
Under a particular $SL(2,R)_L$ transformation with $\ga = 0$ the
string becomes to move with the different velocity
\begin{equation}
x' = \frac{(1 + v)\al^2 - (1 - v)}{(1 + v)\al^2 + 1 - v}t' +
\frac{(1 - v )\al\be}{(1 + v)\al^2 + 1 - v}.
\end{equation}

Now we apply the $SL(2,R)_L$ transformation to the accelerating string
solution (\ref{ex}) which is figured out by the expanding semicircle 
$x^2 + z^2 = t^2 + b^2$ for $0 \le z$ in the $(x,z)$ plane. 
The first relation in 
(\ref{sl}) gives $x + t$ expressed in terms of $x' + t'$, which is 
substituted into the third relation in (\ref{sl}) to obtain $z$ 
expressed in terms of $x' + t'$ and $z'$.
We combine $x - t = (b^2 - z^2)/(x + t)$  with the second relation in
(\ref{sl}) to derive
a third-order equation for $x'$ which has two solutions 
\begin{eqnarray}
x' &=& - t' + \frac{\al}{\ga},
\label{ebg}  \\
\left( x' + \frac{b^2\ga - \be}{2\de}\right)^2 &=&  
\left( t' - \frac{b^2\ga + \be}{2\de}\right)^2 + \frac{b^2}{\de^2}
- {z'}^2.
\label{be}\end{eqnarray}
The former (\ref{ebg}) corresponds to (\ref{xts}) with $A = 1$ so that 
it does not satisfy the string equation, while
the latter (\ref{be}) shows the expanding string where the acceleration
of a quark-antiquark pair changes
from $1/b$ to  $\de/b$.  In a particular $\ga = 0$ case the transformed
 string configuration becomes a second-order equation for $x'$ which is 
also given by (\ref{be}) with $\ga = 0$. In a particular $\de = 0$ case 
the accelerating string solution is transformed to 
a moving string solution with constant velocity
\begin{equation}
x' = \frac{\be^2 - b^2}{\be^2 +  b^2}t' - \frac{b^2 \al\be}{\be^2 +  b^2}
\end{equation}
and the expression (\ref{ebg}) which does not obey the string equation.

When the $SL(2,R)_R$ transformation (\ref{slr}) is applied to
the moving string with constant velocity (\ref{cm}) and
the expanding string (\ref{ex}), the generally 
mapped  string configurations
are described  by (\ref{al}) with $t', v$
replaced by $-t', -v$ and (\ref{ebg}), (\ref{be}) 
with $t'$ replaced by $-t'$
respectively.

Here we consider the special conformal transformation of the Poincare
$AdS_3$ spacetime coordinates $z, x^{\mu}=(t,x)$ 
\begin{equation}
{x^{\mu}}' = \frac{x^{\mu} + a^{\mu}( z^2 + x^2 )}{1 + 2a\cdot x
+ a^2( z^2 + x^2 )}, \hspace{1cm} z' = \frac{z}{1 + 2a\cdot x
+ a^2( z^2 + x^2 )},
\end{equation}
which was studied for the circular Wilson loop \cite{BCF}.
The two cases with $a^{\mu} = (-a,a)$ and $a^{\mu} = (a,a)$
coincide with the particular $SL(2,R)_L$ 
and $SL(2,R)_R$ transformations with $\al = 1, \be = 0, \ga =2a, 
\de = 1$ respectively.

Let us make a special conformal transformation with $a^{\mu} = (0,1/l)$
for the straight string moving with constant velocity $v$ (\ref{cm})
to have  
\begin{equation}
t' = \frac{t}{P}, \hspace{1cm} x' = 
\frac{1}{P}\left( vt + \frac{z^2 - (1-v^2 )t^2}{l} \right),
\hspace{1cm} z' = \frac{z}{P},
\end{equation}
where $P$ is expressed in terms of $t', z'$ through
\begin{equation}
P = 1 + \frac{2vt'}{l}P - \frac{(1 - v^2){t'}^2 - {z'}^2}{l^2}P^2.
\end{equation}
These expressions lead to the expanding string with acceleration
$2/l\sqrt{1-v^2}$
\begin{equation}
x' = \frac{l}{2} \pm \sqrt{\left( t' - \frac{vl}{2}\right)^2 + 
\frac{(1-v^2)l^2}{4} - {z'}^2 }.
\end{equation}
On the other hand a special conformal transformation 
with $a^{\mu} = (1/l,0)$ generates  the expanding string
with acceleration $2v/l\sqrt{1-v^2}$ in the same way
\begin{equation}
x' = -\frac{l}{2v} \pm \sqrt{\left( t' + \frac{l}{2}\right)^2 + 
\frac{(1-v^2)l^2}{4v^2} - {z'}^2 }.
\end{equation}

For the expanding string (\ref{ex}) with acceleration $1/b$ we perform a
special conformal transformation with $a^{\mu} = (0,1/l)$ to obtain
\begin{equation}
t' = \frac{t}{P}, \hspace{1cm} x' = \frac{x + b^2/l}{P}, 
 \hspace{1cm} z' = \frac{z}{P}
\label{tp}\end{equation}
with
\begin{eqnarray}
P &=& 1 + \frac{b^2}{l^2} + \frac{2x}{l} \nonumber \\
&=& 1 + \frac{b^2}{l^2} \pm \frac{2}{l}
\sqrt{b^2 + ( {t'}^2 - {z'}^2 )P^2}.
\label{pl}\end{eqnarray}
The second relation in (\ref{tp}) together with (\ref{pl}) leads to
\begin{equation}
x' = \frac{l}{2} + \left( \frac{b^2}{2l} - \frac{l}{2} \right) 
\frac{1}{P},
\end{equation}
which becomes through the solution $P$ of (\ref{pl}) to be
\begin{equation}
x' = \frac{b^2}{l(b^2/l^2 - 1)} \pm \sqrt{ {t'}^2 + \frac{b^2}
{( b^2/l^2 - 1 )^2} - {z'}^2 }
\end{equation}
for $b/l \neq \pm 1$. Thus the magnitude of 
acceleration changes from $1/b$
to $|b^2/l^2 - 1|/b$.
In particular $ l = \pm b$ cases the expanding string turns back to a 
static straight string located at $x' = \pm b/2$ stretching from the AdS
boundary to the Poincare horizon, which is dual to a static isolated
quark. 
 
The other special conformal transformation with $a^{\mu} = (1/l,0)$
is applied to the expanding string (\ref{ex}) as
\begin{equation}
t' = \frac{t + b^2/l}{P}, \hspace{1cm} x' = \pm
\frac{\sqrt{t^2 + b^2 - z^2}}{P},  \hspace{1cm} z' = \frac{z}{P}
\label{tz}\end{equation}
with $P = 1 -2t/l - b^2 /l^2$. The first relation in (\ref{tz}) reads 
\begin{equation}
t = \frac{(1 - b^2 /l^2)t' - b^2 /l}{1 + 2t'/l},
\label{ta}\end{equation}
which is substituted into the third relation in (\ref{tz}) to generate
\begin{equation}
z = \frac{1 + b^2 /l^2}{1 + 2t'/l}z'.
\label{si}\end{equation}
Combining (\ref{ta}), (\ref{si}) with the second relation in (\ref{tz})
we have again  the expanding string configuration with
acceleration $(1 + b^2 /l^2)/b$
\begin{equation}
x' = \pm \sqrt{\left( t' + \frac{b^2}{l( 1 + b^2 /l^2)} \right)^2 + 
\frac{b^2}{( 1 + b^2 /l^2)^2} - {z'}^2 }.
\end{equation}

Here we restore the AdS radius $R$  to express the following relations
between the Poincare coordinates in $AdS_3$ and the embedding coordinates
$X^M\; (M = -1, 0, 1, 2 )$ on which the conformal SO(2,2) transformation
is acting linearly
\begin{eqnarray}
X^{\mu} &=& \frac{x^{\mu}}{z}R, \; (\mu =0, 1), \nonumber \\
X^{-1} &=& \frac{R^2 + z^2 + x_{\mu}x^{\mu} }{2z}, \hspace{1cm}
X^2 = \frac{R^2 - z^2 - x_{\mu}x^{\mu} }{2z}, \nonumber \\
- R^2 &=& - (X^{-1})^2 - (X^0)^2 + (X^1)^2 + (X^2)^2 .
\end{eqnarray}

For the moving string  with  constant velocity $v$, which is described
by $X^1 = v X^0$, we perform one conformal SO(2,2) transformation 
\begin{equation} 
{X^{-1}}' = - X^0, \;\; {X^0}' = X^{-1}, \;\; {X^1}' = X^1, \;\;
{X^2}' = X^2,
\label{os}\end{equation}
which interchanges $X^{-1}$ and $X^0$. The transformed configuration
is specified by ${X^1}' = - v {X^{-1}}'$ that is expressed in terms of
the Poincare coordinates as
\begin{equation}
\left(x' + \frac{R}{v}\right)^2 = {t'}^2 + 
\frac{1 - v^2}{v^2}R^2 - {z'}^2,
\end{equation}
which represents the expanding string with acceleration $v/R\sqrt{1-v^2}$.

The other conformal SO(2,2) transformation defined as the interchange 
between $X^1$ and $X^2$
\begin{equation}
{X^{-1}}' = X^{-1}, \;\; {X^0}' = X^{0}, \;\; {X^1}' = - X^2, \;\;
{X^2}' = X^1
\label{so}\end{equation}
produces ${X^2}' = v {X^0}'$ which becomes
\begin{equation}
{x'}^2 = ( t' - vR )^2 + (1 - v^2)R^2 - {z'}^2.
\end{equation}
Thus the expanding string solution with acceleration
$1/R\sqrt{1-v^2}$ is constructed.

Now performing the conformal SO(2,2) transformation (\ref{os}) for
the expanding string (\ref{ex}) which is expressed as
\begin{equation}
 (X^0)^2 - (X^1)^2 = R^2 - \frac{b^2}{R^2}(X^{-1} + X^2)^2 
\label{xr}\end{equation}
we obtain a curve
\begin{equation}
 ({X^{-1}}')^2 - ({X^1}')^2 = R^2 - \frac{b^2}{R^2}({X^0}' + {X^2}')^2.
\label{cn}\end{equation}
The mapped expression is a polynomial of $x'$ in the fourth order
which is compared with the second-order plynomial of $x$ in (\ref{xr}).
It, however, is expressed in terms of $y \equiv {x'}^2 + {z'}^2$ as
\begin{eqnarray}
\left( 1 + \frac{b^2}{R^2}\right) y^2 &-& 2 \left( {t'}^2 + R^2   
+ \frac{b^2}{R^2}(t' + R)^2 \right)y \nonumber  \\
 +\; ({t'}^2 - R^2)^2 &+&  \frac{b^2}{R^2}(t' + R)^4 = 0
\end{eqnarray}
so that we have two solutions
\begin{eqnarray}
{x'}^2 &=& (t' + R)^2 - {z'}^2, 
\label{xd} \\
{x'}^2 &=& \left(t' + \frac{b^2 - R^2}{b^2 + R^2}R \right)^2 
+ \frac{4b^2R^4}{(b^2 + R^2)^2} - {z'}^2.
\label{rt}\end{eqnarray}
The former (\ref{xd}) corresponds to (\ref{ex}) with $b = 0$ so that
it does not solve the string equation, while
the latter (\ref{rt}) is the expanding string solution with 
acceleration $(b^2 + R^2)/2bR^2$.

The other conformal SO(2,2) transformation (\ref{so}) applied to 
the expanding string solution (\ref{xr}) produces a curve
\begin{equation}
 ({X^0}')^2 - ({X^2}')^2 = R^2 - \frac{b^2}{R^2}({X^{-1}}' - {X^1}')^2,
\label{cn}\end{equation}
which is similarly expressed in terms of $y \equiv {t'}^2 - {z'}^2$  as
\begin{eqnarray}
\left( 1 - \frac{b^2}{R^2}\right) y^2 &-& 2 \left( {x'}^2 + R^2   
- \frac{b^2}{R^2}(x' - R)^2 \right)y \nonumber  \\
 + \;({x'}^2 - R^2)^2 &-&  \frac{b^2}{R^2}(x' - R)^4 = 0. 
\label{yr}\end{eqnarray}
For $ R \neq b$ two solutions are obtained by
\begin{eqnarray}
{t'}^2 - {z'}^2 &=& (x' - R)^2, \label{rb} \\
{t'}^2 - {z'}^2 &=& \frac{1}{1 - b^2/R^2}\left( (x' + R)^2 - 
\frac{b^2}{R^2}(x' - R)^2 \right).
\label{br}\end{eqnarray}
The former (\ref{rb}) also does not obey the string equation,
while the latter (\ref{br}) yields the expanding string
solution with acceleratioin $|b^2 - R^2|/2bR^2$ as shown by
\begin{equation}
\left(x' - \frac{b^2 + R^2}{b^2 - R^2}R \right)^2 = {t'}^2 
+ \frac{4b^2R^4}{(b^2 - R^2)^2} - {z'}^2.
\end{equation}
In a particular $R = b$ case the equation (\ref{yr}) leads to
\begin{equation}
x'\left( ( x' - R)^2 - ( {t'}^2 - {z'}^2 ) \right) = 0,
\end{equation}
whose solutions are given by $x' = 0$ and $x' = R \pm 
\sqrt{{t'}^2 - {z'}^2}$. The former shows the static string
solution, while the latter does not satisfy the string equation.

\section{The accelerating string in the string sigma model
action}

Let us consider a time-dependent open string configuration in $AdS_3$
with the Poincare metric by analyzing the string sigma model action
in the Lorentzian worldsheet coordinates with $a=0,1$
\begin{equation}
S = - \frac{\sqrt{\lambda}}{4\pi}\int d\tau d\sigma \frac{1}{z^2}
(- \pa_at\pa^at + \pa_ax\pa^ax + \pa_az\pa^az ).
\end{equation} 
The off-diagonal Virasoro constraint gives 
\begin{equation}
- \dot{t}t' + \dot{x}x' + \dot{z}z' = 0.
\label{tt}\end{equation}
In this section we use the dot and prime as the derivatives with respect
to $\tau$ and $\sigma$ respectively. 

Here we choose the following ansatz in the factorized form
\begin{equation}
t = t_{\tau}(\tau)f(\sigma), \hspace{1cm} x = x_{\tau}(\tau)f(\sigma), 
\hspace{1cm} z = z(\sigma).
\label{fa}\end{equation}
The off-diagonal Virasoro constraint  (\ref{tt}) reads
\begin{equation}
- \dot{t_{\tau}}t_{\tau} + \dot{x_{\tau}}x_{\tau} = 0,
\end{equation}
which is solved by
\begin{equation}
x_{\tau}^2  - t_{\tau}^2 = \pm N^2
\label{xn}\end{equation}
with an integration constant $N$. 

First we consider the plus case to parameterize $x_{\tau}$ and 
 $t_{\tau}$ in terms of a positive parameter $p$ as
\begin{equation}
x_{\tau} = N\cosh p\tau, \hspace{1cm}  t_{\tau} = N\sinh p\tau
\label{nh}\end{equation}
with $ -\infty < \tau < \infty$. The integration 
constant $N$ is absorbed into
$f(\sigma)$ so that $N$ can be set to unity.

The diagonal Virasoro constraint gives
\begin{equation}
{f'}^2 - p^2f^2 +{z'}^2 = 0.
\label{vi}\end{equation}
Substituting the ansatz (\ref{fa}) with (\ref{nh}) into the equations of
motion for $t$ and $x$
\begin{equation}
\pa_{\tau} \left( \frac{\dot{t}}{z^2} \right) - 
\pa_{\sigma} \left( \frac{t'}{z^2} \right) = 0, \;\;
\pa_{\tau} \left( \frac{\dot{x}}{z^2} \right) - 
\pa_{\sigma} \left( \frac{x'}{z^2} \right) = 0
\label{ext}\end{equation}
we have an identical equation 
\begin{equation}
f'' = \frac{2z'}{z}f' + p^2f.
\label{pf}\end{equation}
The equation of motion for $z$
\begin{equation}
\pa_{\tau} \left( \frac{\dot{z}}{z^2} \right) - 
\pa_{\sigma} \left( \frac{z'}{z^2} \right) = \frac{1}{z^3}\left(
(\pa_az)^2 -  (\pa_at)^2 + (\pa_ax)^2 \right)
\label{ez}\end{equation}  
 turns out to be
\begin{equation}
zz'' = {z'}^2 - {f'}^2 - p^2f^2.
\label{zz}\end{equation}
We sum (\ref{zz}) and (\ref{pf}) multiplied by $f$ to derive
a differential equation
\begin{equation}
\pa_{\sigma}^2( z^2 + f^2  ) = \frac{2z'}{z} \pa_{\sigma}(  z^2 + f^2 ).
\end{equation}

Here we consider a simple solution 
\begin{equation}
 z^2 + f^2 = b^2
\label{fb}\end{equation}
with a constant positive parameter $b$, which implies $0 \le z \le b$.
We substitute $f = \pm\sqrt{b^2 - z^2}$ of (\ref{fb}) into (\ref{vi})
to obtain an equation for $z$
\begin{equation}
b^2{z'}^2 - p^2 ( b^2 - z^2 )^2 = 0.
\label{bp}\end{equation}
Owing to $0 \le z \le b$ the solution of (\ref{bp}) is expressed as
\begin{equation}
z = b \tanh p\sigma,
\label{zh}\end{equation}
where we take the range of $\sigma$ as $0 \le \sigma < \infty$.
In the region $0 \le z \le b$ we have a string profile expressed by 
(\ref{zh}) and
\begin{equation}
x = \pm \frac{b\cosh p\tau}{\cosh p\sigma}, \hspace{1cm}
t = \pm \frac{b\sinh p\tau}{\cosh p\sigma}.
\label{sx}\end{equation}
Hereafter the plus and minus  
solutions are called by the I and II solutions
respectively. These I and II solutions are confirmed to satisfy the
string equations (\ref{ext}) and (\ref{ez}).
Eliminating the dependences of the worldsheet coordinates we reproduce
the accelerating string solution (\ref{ex}) 
for the region $0 \le z \le b$.

Alternatively we replace $\sigma$ by $z$ through (\ref{zh}) to have 
\begin{equation}
x = \pm \sqrt{b^2 - z^2}\cosh p\tau, \hspace{1cm}
t = \pm \sqrt{b^2 - z^2}\sinh p\tau.
\label{sth}\end{equation}
If we choose $p = 1/b$, then in the AdS boundary $z = 0$, that is, 
$\sigma = 0$ the solutions (\ref{sth}) become
$x = \pm b\cosh \tau/b$ and $t = \pm b\sinh \tau/b$, which yield
$x = \pm \sqrt{t^2 + b^2}$ and represent the accelerating quark and
antiquark trajectories for plus and minus signs respectively with 
proper time $\tau$ and proper acceleration $1/b$.
The expressions of (\ref{zh}) and (\ref{sth}) with $p = 1/b$
agree with ones in ref. \cite{BX} which are described as
a static solution in the generalized Rindler spacetime that is derived
from the AdS spacetime by a coordinate transformation.
The string  in the generalized Rindler spacetime
 was analyzed in \cite{CCG,HKK} where the thermodynamics associated with
the worldsheet horizon which has the Unruh temperature is
further  studied. 

For fixed $\sigma$, that is, fixed $z$ in the I solution the limit 
$\tau = -\infty$ leads to $t = -\infty,\: x = \infty$ and $\tau = 0$ gives
 $t = 0, \; x = b/\cosh(\sigma/b) = \sqrt{b^2 - z^2}$ 
that is the position of the
string bit at depth $z$ and at time $t = 0$.
The limit $\tau = \infty$ leads to $t = \infty,\: x = \infty$, and
 there is a restriction $|t|< x$ for each $\tau$ through
$t/x = \tanh \tau/b$. On the other hand 
in the II solution, $\tau = \infty$ corresponds to  $t = -\infty,\: 
x = -\infty$, $\tau = 0$ to $t = 0, x = -\sqrt{b^2 - z^2}$ and
$\tau = -\infty$ to $t = \infty,\: x = -\infty$, which imply
 $x < -|t|$ for each $\tau$.

For fixed $t$ the I string extends from the quark location 
$(x,z) = (\sqrt{t^2 + b^2},0)$ specified by $\sigma = 0$ 
to $(x,z) = (|t|,b)$ in an arc, while the II string extends from the 
antiquark location $(x,z) = (-\sqrt{t^2 + b^2},0)$ to
$(x,z) = (-|t|,b)$ similarly.

Here let us consider the minus case for (\ref{xn}) with $N = 1$
and represent $x_{\tau}$ and $t_{\tau}$ as
\begin{equation}
x_{\tau} = \sinh p\tau, \hspace{1cm}  t_{\tau} = \cosh p\tau.
\end{equation}
The diagonal Virasoro constraint yields
\begin{equation}
-{f'}^2 + p^2f^2 + {z'}^2 = 0.
\label{ft}\end{equation}
In this case the equations of motion for $t$ and $x$ lead to
the same equation as (\ref{pf}), however, the equation of motion for $z$
gives
\begin{equation}
zz'' = {z'}^2 + {f'}^2 + p^2f^2.
\label{fp}\end{equation}

Combining together we derive 
\begin{equation}
\pa_{\sigma}^2( z^2 - f^2  ) = \frac{2z'}{z} \pa_{\sigma}( z^2 - f^2 ),
\end{equation}
which has two simple solutions $z^2 - f^2 = \pm b^2$, where
the upper sign case has  a restriction $b \le z$.
The equation (\ref{ft}) with $f^2 = z^2 \mp b^2$ can be expressed as
\begin{equation} 
\mp \frac{b^2 {z'}^2}{z^2 \mp b^2} + p^2(z^2 \mp b^2) = 0.
\end{equation}
This equation of $z$ for the lower sign has no real solution, while
 for the upper sign it has the following solution
\begin{equation}
z = \frac{b}{\tanh p\sigma},
\label{zs}\end{equation}
which yields
\begin{equation}
x = \pm \frac{b\sinh p\tau}{\sinh p\sigma}, \hspace{1cm}
t = \pm \frac{b\cosh p\tau}{\sinh p\sigma}
\label{xs}\end{equation}
with $0 \le \sigma  < \infty,\; -\infty <\tau < \infty$.
To the plus and minus solutions we call the III and IV solutions. 
These solutions are  expressed as 
\begin{equation}
x = \pm \sqrt{z^2 - b^2}\sinh p\tau, \hspace{1cm}
t = \pm \sqrt{z^2 - b^2}\cosh p\tau,
\label{ts}\end{equation}
which also reproduce the accelerating string solution (\ref{ex})
through the elimination of $\tau$ for the region $b \le z \le 
\sqrt{t^2 + b^2}$.

In the IV solution of (\ref{xs}) and (\ref{ts}) with $p = 1/b$ for fixed
$\tau$, $t$ is negative and changes from $t = - \infty$ at $\sigma = 0$ to
$t = 0$ at $\sigma = \infty$, where the roles of $\tau$ and $\sigma$
are exchanged in comparison with the I and II solutions.
 Owing to $x/t = \tanh\tau /b$, $x$ varies such
that $x = t = -|t|$ at $\tau = \infty, \; x = 0$ at $\tau = 0$ and 
$x = -t = |t|$ at $\tau = - \infty$. 
At fixed $t < 0$ the IV string extends from $(x,z) = (-|t|,b)$ to 
$(x,z) = (|t|,b)$ in an arc and shrinks to zero at $t = 0$

On the other hand in the III solution 
for fixed $\tau$, $t$ is positive and
changes from $t = 0$ at $\sigma = \infty$ to $t = \infty$ at $\sigma = 0$.
 Owing to $x/t = \tanh\tau /b$, $x$ varies such
that $x = - t$ at $\tau = -\infty, \; x = 0$ at $\tau = 0$ and 
$x = t$ at $\tau = \infty$. At fixed $t, \; z$ is described by $z =
b(1 + t^2/(b\cosh \tau/b)^2)^{1/2}$ so that $z$ becomes $z = b$ 
at $\tau = \pm \infty$ that implies $\sigma = \infty$ 
for $t$ to be fixed, and
$z = \sqrt{t^2 + b^2}$ at $\tau = 0$. Thus at 
$t = 0$ the III string starts
as a point at $(x,z) = (0,b)$ and at fixed $t >0$
 extends from $(x,z) = (-t,b)$ to 
$(x,z) = (t,b)$  through $(x,z) = (0,\sqrt{t^2 + b^2}) = 
(0,b/\tanh \sigma/b)$.

Now we calculate the induced metric on the string surface, (\ref{zh}) and
(\ref{sx}) in the region $0 \le z \le b$ 
to obtain a conformally flat expression
\begin{equation}
ds_{ws}^2 = \frac{1}{b^2\sinh^2(\sigma/b)} (- d\tau^2 + d\sigma^2 ),
\label{ws}\end{equation}
which has a horizon at $\sigma = \infty$ on the worldsheet that 
yields $z = b$ as a dividing line in the bulk spacetime.
The induced metric on the string surface, (\ref{zs}) and
(\ref{xs}) in the region $b \le z$ is described by a different 
expression
\begin{equation}
ds_{ws}^2 = \frac{1}{b^2\cosh^2(\sigma/b)} ( d\tau^2 - d\sigma^2 ),
\label{wc}\end{equation}
which has also a horizon at $\sigma = \infty$ on the worldsheet that 
corresponds to $z = b$.

If we make a coordinate transformation from $\sigma$ to $z$ using 
(\ref{zh}) and (\ref{zs}) to rewrite (\ref{ws}) and 
(\ref{wc}) in terms of $\tau$ and $z$ respectively, we obtain a 
single expression
\begin{equation}
ds_{ws}^2 = \frac{1}{z^2} \left( -\left(1 - \frac{z^2}{b^2} \right)
d\tau^2 + \frac{dz^2}{1 - z^2/b^2} \right),
\end{equation}
where there is a horizon at $z = b$.
In the interior region $b \le z$ the roles of $\tau$
and $z$ are exchanged such that $\tau$ becomes a spacelike coordinate
 and $z$ becomes a timelike coordinate, which corresponds to the exchange
of  the roles of $\tau$ and $\sigma$ between (\ref{ws}) and (\ref{wc}).

Combining the string solutions I, II, III and IV derived from the
string sigma model action, we have the following picture.
In the early time specified by $t < 0$, the right string I and the
left string II staying in the exterior region $0 \le z \le b$
are decelerated and connected at $z = b$ by the middle string IV 
staying in the interior region $b \le z$. At $t = 0$ 
the interior string IV shrinks
to zero and the two exterior strings I and II stop and directly 
touch at $z= b$. In the late time $t > 0$, the exterior strings I and II
return back and are accelerated in opposite directions, where the
two exterior strings are connected by the interior string III. 

\section{Conclusion}

For the open string in the Poincare $AdS_3$ spacetime we have used the
Nambu-Goto action in the static gauge to make an ansatz in a 
square root expression characterized by three parameters for
the string profile. We have observed that if three parameters are 
appropriately chosen, there appear two open string solutions
in a complementary pair, the string solution
associated with the one-cusp Wilson loop \cite{MK}
and the expanding string solution associated with a uniformly 
 accelerating quark-antiquark pair \cite{BX}.

We have constructed the expanding string solution by applying 
the $SL(2,R)_L \times SL(2,R)_R$ isometry 
transfomations, the special conformal
 transformations and the conformal SO(2,2) transformations to a simple
moving string solution  dual to  a moving quark with constant velocity.
 We have  demonstrated that under the three kinds of 
transformations the expanding string solution is usually mapped to the
same expanding string with the different acceleration.
It has been observed that some particular transformations make
the expanding string solution change back to the moving string
solution with constant velocity or the static string solution.

Based on the string sigma  model action 
we have made an ansatz for the open
string profile in the factorized form and constructed two kinds of 
string solutions,  the exteror strings and the interior 
strings that stay in the two different bulk spacetime regions. 
We have observed that on each string worldsheet 
there appear the horizon which is associated with the dividing line 
which separates the two different bulk spacetime regions.  
We have demonstrated that the expanding string solution is constructed by
connecting two separated exterior strings with one interior string.

\end{document}